\documentclass[10pt]{article}

\usepackage{amsmath,amssymb}
\usepackage[ansinew]{inputenc}
\usepackage{graphicx}
\usepackage{color}
\usepackage{multicol}
\allowdisplaybreaks

\newcommand{\dd}[2]{\frac{\mathrm{d} #1}{\mathrm{d} #2}}
\newcommand{\pdl}[2]{\frac{\partial #1}{\partial #2}}
\newcommand{\pds}[2]{\partial #1 / \partial #2}

\newcommand{\p}{\partial}

\newcommand{\todo}[1][\null]{\ensuremath{\clubsuit}}
\newcommand{\const}{\mathop{\rm const}\nolimits}

\newcommand{\vv}{\mathbf{v}}
\newcommand{\ww}{\mathbf{w}}

\newcommand{\ve}{\varepsilon}

\newcommand{\Ad}{\mathrm{Ad}}

\newcommand{\DDD}{\mathcal{D}}

\begin{document}


\par\noindent {\LARGE\bf
Symmetry Analysis of Barotropic Potential \\Vorticity Equation
\par}
{\vspace{4mm}\par\noindent {\bf Alexander Bihlo~$^\dag$ and Roman O. Popovych~$^{\dag,\ddag}$
} \par\vspace{2mm}\par}

{\vspace{2mm}\par\noindent {\it
$^\dag$~Faculty of Mathematics, University of Vienna, Nordbergstra{\ss}e 15, A-1090 Vienna, Austria\\
}}
{\noindent \vspace{2mm}{\it
$\phantom{^\dag}$~\textup{E-mail}: alexander.bihlo@univie.ac.at
}\par}

{\vspace{2mm}\par\noindent {\it
$^{\ddag}$~Institute of Mathematics of NAS of Ukraine, 3 Tereshchenkivska Str., 01601 Kyiv, Ukraine\\
}}
{\noindent \vspace{2mm}{\it
$\phantom{^\dag}$~\textup{E-mail}: rop@imath.kiev.ua
}\par}

{\vspace{7mm}\par\noindent\hspace*{8mm}\parbox{140mm}{\small
Recently F.~Huang [Commun. Theor. Phys. \textbf{42} (2004) 903] and X.~Tang and P.K.~Shukla [Commun. Theor. Phys. \textbf{49} (2008) 229] 
investigated symmetry properties of the barotropic potential vorticity equation without forcing and dissipation on the beta-plane. This equation 
is governed by two dimensionless parameters, $F$ and $\beta$, representing the ratio of the characteristic length scale to the Rossby radius of 
deformation and the variation of earth' angular rotation, respectively. In the present paper it is shown that in the case $F\ne 0$ there exists 
a well-defined point transformation to set $\beta = 0$. The classification of one- and two-dimensional Lie subalgebras of the Lie symmetry 
algebra of the potential vorticity equation is given for the parameter combination $F\ne 0$ and $\beta = 0$. Based upon this classification, 
distinct classes of group-invariant solutions is obtained and extended to the case~$\beta \ne 0$.
\newline\newline
\noindent\textbf{PACS numbers:} 47.35.-i, 47.32.-y, 02.20.Sv\\
\noindent\textbf{Key words:} potential vorticity equation, Lie symmetries, classification of subalgebras, exact solutions

}\par\vspace{7mm}}


\begin{multicols}{2}

\section{Introduction}

There is a long history in dynamic meteorology to use simplified models of the complete set of hydro-thermodynamical equations to gain insides 
in the different processes characterising the various structures and pattern occurring in the atmosphere. One of the most classical models in 
atmospheric science is the barotropic (potential) vorticity equation. It has been successfully used both for theoretical considerations  
\cite{ross39Ay,char79Ay} and practical numerical weather predictions \cite{char50Ay} since it is capable of describing some prominent features 
of mid-latitude weather phenomena such as the well-known Rossby waves and blocking regimes. In nondimensional form it reads~\cite{pedl87Ay}
\begin{align}\label{vort}
\begin{split}
   & \pdl{\zeta}{t} - F\pdl{\psi}{t} + J(\psi,\zeta) + \beta\pdl{\psi}{x} = 0, \\
   & \zeta = \pdl{^2\psi}{x^2}+\pdl{^2\psi}{y^2}.
\end{split} 
\end{align}
Here $\psi$ is the stream function, $\zeta$ the vorticity, $J(a,b) = \pds{a}{x}\pds{b}{y}-\pds{a}{y}\pds{b}{x}$ is the Jacobian and $F$ and 
$\beta$ represent the ratio of the characteristic length scale to the Rossby radius of deformation and a parameter describing the variation of 
earth' angular rotation, respectively.

\section{The Lie symmetries}

The symmetries of (\ref{vort}) have first been investigated in \cite{blen91Ay} and were applied to construct new solutions from known ones.
Later, they have been studied by \cite{huan04Ay} and \cite{tang08Ay}, but without reference to the group 
classification problem. This problem arises since different values of $F$ and $\beta$ lead to different symmetry properties of (\ref{vort}), 
which in turn characterise different physical properties of model (\ref{vort}). However, as shown in this paper, there are only three essential 
combinations of the values of these two parameters, given by $F=0$, $\beta=0$; $F=0$, $\beta\ne 0$ and $F\ne 0$. The first combination leads to 
the usual vorticity form of Euler's equation which is the issue of e.g. \cite{andr98Ay}. This combination is of particular interest, since it 
gives rise to a new symmetry \cite{berk63Ay} (a so-called potential symmetry) which is not present in the velocity form of Euler's equation. The 
second combination of parameters was discussed in \cite{bihl07Ay,ibra95Ay}. It is the usual barotropic vorticity equation. The parameter~$\beta$ 
can be set to~1 by scaling and/or changing signs of variables. Note that for both combinations the associated Lie invariance algebras are 
infinite dimensional.

Now we show that if $F\ne 0$, we can always set $\beta=0$. (Then the nonvanishing parameter~$F$ can be scaled to~$\pm1$.) For this purpose, we 
recompute the symmetries of (\ref{vort}) for the case $F\ne 0$ and $\beta$ arbitrary. This is done upon using the computer algebra packages MuLie 
\cite{head93Ay} and DESOLV \cite{carm00Ay}. For this combination, equation (\ref{vort}) admits the six-dimensional Lie symmetry algebra 
$\mathfrak a_\beta$ generated by the operators
\begin{align}\label{sym1}
\begin{split}
    \DDD &= t\p_t - \frac{\beta}{F}t\p_x - \left(\psi - \frac{\beta}{F}y\right)\p_\psi, \\
    \vv_r &= -y\p_x + \left(x+\frac{\beta}{F}t\right)\left(\p_y + \frac{\beta}{F}\p_\psi\right), \\
    \vv_t &= \p_t, \\
    \vv_x &= \p_x, \\
    \vv_y &= \p_y, \\
    \vv_\psi &= \p_\psi.
\end{split}
\end{align}
This algebra is not singular in $\beta$ and consequently it also includes the case $\beta=0$. Moreover, computing the symmetries for the case 
 $\beta=0$ explicitly, we obtain the same algebra (\ref{sym1}) with $\beta=0$. Hence
\begin{align}\label{sym2}
\begin{split}
    \DDD &= t\p_t - \psi \p_\psi,   \\
    \vv_r &= -y\p_x + x\p_y, \\
    \vv_t &= \p_t, \\ 
    \vv_x &= \p_x, \\
    \vv_y &= \p_y, \\
    \vv_\psi &= \p_\psi
\end{split}
\end{align}
are the generators of the Lie symmetry algebra $\mathfrak a_0$. The physical importance of these generators is the following: $\DDD$ generates 
simultaneous scaling in $t$ and $\psi$, $\vv_r$ is the rotation operator in the $(x,y)$-plane, $\vv_t$, $\vv_x$, $\vv_y$ and $\vv_\psi$ are the 
infinitesimal generators of translations in $t$, $x$, $y$ and $\psi$, respectively. The nonzero commutation relations between basis elements 
\eqref{sym2} are exhausted by 
\[
\begin{split}
&[\vv_t,\DDD]=\vv_t, \quad [\vv_\psi,\DDD]=-\vv_\psi,\\
&[\vv_x,\vv_r]=\vv_y, \quad [\vv_y,\vv_r]=-\vv_x. 
\end{split}
\]
Therefore, the algebra $\mathfrak a_0$ has a simple structure. 
It is a solvable Lie algebra and can be represented as the direct sum $\mathfrak g_{3.4}^{-1}\oplus\mathfrak e(2)$, 
where $\mathfrak e(2)=\langle\vv_x,\vv_y,\vv_r\rangle$ is the Euclidean algebra in the $(x,y)$-plane 
and $\mathfrak g_{3.4}^{-1}=\langle\vv_t,\vv_\psi,\DDD\rangle$ is a three-dimensional almost Abelian Lie algebra 
from Mubarakzyanov's classification of low dimensional Lie algebras \cite{muba63Ay}. 

It is straightforward to show that the Lie algebra (\ref{sym1}) maps to the Lie algebra (\ref{sym2}) under the transformation given by
\begin{equation}\label{trans}
\begin{split}
    \tilde t &= t, \quad \tilde x = x + \frac{\beta}{F}t, \quad \tilde y = y, \\ 
    \tilde \psi &= \psi - \frac{\beta}{F}y.
\end{split}
\end{equation}
This transformation also maps (\ref{vort}) to the same equation with $\beta=0$. That is, (\ref{trans}) is an equivalence transformation for the 
class of equations of the form (\ref{vort}) with $F\ne 0$. Hence the $\beta$-term can be neglected under symmetry analysis. Every solution of 
the equation with $\beta=0$ can be extended to a solution with $\beta \ne 0$ by means of the transformation (\ref{trans}). 

We also note that the maximal (infinite dimensional) Lie symmetry algebras in the cases $F=0$, $\beta=0$ and $F=0$, $\beta\ne 0$ are neither 
isomorphic to each other nor isomorphic to the (finite dimensional) algebra $\mathfrak{a}_0$. Consequently, it is not possible to find point 
transformations that relate the corresponding PDEs to each other. All the transformations used for the reductions of the parameters $F$ and 
$\beta$ belong to the equivalence group of  class~(\ref{vort}). The group classification list for class~(\ref{vort}) is therefore exhausted by 
the three inequivalent cases $F=0$, $\beta=0$; $F=0$, $\beta=1$ and $F=\pm1$, $\beta=0$.

\section{Classification of subalgebras}

Classification of subgroups of Lie symmetry groups of differential equations is an essential part in the study of these equations. This is since 
classification allows for an efficient computation of group-invariant solutions, without the possibility of an occurrence of equivalent 
solutions. Classifying subgroups may further lead to the construction of simple ans\"atze for the corresponding equivalence classes of reduced 
differential equations. Thereby, the classification also provides an important step for further investigations of properties of these reduced 
equations.

The classification of subgroups of symmetry groups is usually done by the classification of the associated Lie subalgebras with respect to the 
adjoint representation \cite{olve86Ay,ovsi82Ay}. The potential vorticity equation (\ref{vort}) is a (2+1) model and thus Lie reductions up to an 
ordinary differential equations require the classification of one- and two-dimensional subalgebras. An exhaustive classification of subalgebras 
exists only for Lie algebras up to dimension four \cite{pate77Ay}. Thus we need to classify subalgebras of $\mathfrak a_0$. This problem is not 
difficult because the algebra $\mathfrak a_0$ has a simple solvable structure.

The adjoint representation of a Lie group on it's Lie algebra is given as the Lie series
\begin{align}
  \ww(\ve) &= \Ad(e^{\ve\vv})\ww_0 := \sum_{n=0}^{\infty}\frac{\ve^n}{n!}\{\vv^n,\ww_0 \},
\end{align}
with $\{ \cdot,\cdot\}$ being defined recursively: 
\begin{align*}
\{\vv^0,\ww_0\} &:= \ww_0,\\
\{\vv^n,\ww_0\} &:= (-1)^n[\vv,\{\vv^{n-1},\ww_0\}]. 
\end{align*}
Alternatively,~the adjoint representation can also be calculated by integrating the initial value problem
\[
    \dd{\ww(\ve)}{\ve} = [\ww(\ve),\vv], \quad   \ww(0) = \ww_0.
\]
For basis elements (\ref{sym2}) of the algebra $\mathfrak a_0$ we obtain the following nontrivial adjoint actions
\begin{align*}
    \Ad(e^{\ve\vv_t})\DDD &= \DDD - \ve\vv_t, \\[.5ex]
    \Ad(e^{\ve\vv_\psi})\DDD &= \DDD + \ve\vv_\psi, \\[.5ex]
    \Ad(e^{\ve\DDD})\vv_t &= e^{-\ve}\vv_t, \\[.5ex]
    \Ad(e^{\ve\DDD})\vv_\psi &= e^{\ve}\vv_\psi, \\[.5ex]
    \Ad(e^{\ve\vv_x})\vv_r &= \vv_r - \ve\vv_y, \\[.5ex]
    \Ad(e^{\ve\vv_y})\vv_r &= \vv_r + \ve\vv_x, \\[.5ex]
    \Ad(e^{\ve\vv_r})\vv_x &=\phantom{{}-{}} \vv_x\cos\ve + \vv_y\sin\ve, \\[.5ex]
    \Ad(e^{\ve\vv_r})\vv_y &= {}-\vv_x\sin\ve+ \vv_y\cos\ve .
\end{align*}

\subsection{One-dimensional subalgebras}

The classification of one-dimensional subalgebras of the whole symmetry algebra (\ref{sym2}) is done by an inductive approach \cite{olve86Ay}: 
We start with the most general infinitesimal generator,
\[
    \vv = a_1\DDD + a_2\vv_r + a_3\vv_t + a_4\vv_x + a_5\vv_y + a_6\vv_\psi,
\]
and simplify it as much as possible by means of adjoint actions. Depending on the respective values of the coefficients $a_i$, $i=1,\dots,6$, we 
find the following list of inequivalent one-dimensional subalgebras of (\ref{sym2}):
\begin{align}\label{one}
\begin{split}
  1.\,\, a_1&\ne 0, a_2 \ne 0\colon \\[.5ex]
  &\langle \DDD+a\vv_r \rangle. \\[.5ex]
  2.\,\, a_1&\ne 0, a_2=0, a_4\ne 0\colon \\[.5ex]
  &\langle\DDD+a\vv_x \rangle.  \\[.5ex]
  3.\,\, a_1&=0, a_2\ne 0, a_3\ne 0\colon \\[.5ex]
   &\langle\vv_r \pm \vv_t + a\vv_\psi \rangle.  \\[.5ex]
  4.\,\, a_1&=a_3=0, a_2\ne 0\colon \\[.5ex]
   &\langle\vv_r + c\vv_\psi \rangle.  \\[.5ex]
  5.\,\, a_1&=a_2=0, a_3\ne 0,a_4\ne 0\colon \\[.5ex]
    &\langle\vv_t + a\vv_x + c\vv_\psi \rangle.  \\[.5ex]
  6.\,\, a_1&=a_2=a_3=0,a_4\ne 0\colon \\[.5ex]
   &\langle\vv_x + c\vv_\psi \rangle.  \\[.5ex]
  7.\,\, a_1&=a_2=a_3=a_4=a_5=0\colon \\[.5ex]
   &\langle\vv_\psi \rangle.
\end{split}  
\end{align}where $c\in\{-1,0,1\}$ and $a, a_i \in \mathbb{R}$. In case 5 we can additionally set $a\in\{-1,0,1\}$ if $c=0$.

\subsection{Two-dimensional subalgabras}

The classification procedure of two-dimensional subalgebras follows in essential the same way as the one-dimensional case. The two most general 
linearly independent infinitesimal generators
\begin{align*}
    \vv^1 &= a_1^1\DDD + a_2^1\vv_r + a_3^1\vv_t + a_4^1\vv_x + a_5^1\vv_y + a_6^1\vv_\psi, \\
    \vv^2 &= a_1^2\DDD + a_2^2\vv_r + a_3^2\vv_t + a_4^2\vv_x + a_5^2\vv_y + a_6^2\vv_\psi
\end{align*}
are simultaneously subjected to the adjoint actions and nonsingular linear combining under some assumptions on the coefficients $a_i^j$, 
$i=1,\dots,6$, $j=1,2$. Moreover, the required closure property of the subalgebra $\langle\vv^1,\vv^2\rangle$ with respect to the Lie bracket 
(i.e.\ $[\vv^1,\vv^2]\in\langle\vv^1,\vv^2\rangle$) eventually places further restrictions on the coefficients. By applying this technique, we 
find a set of inequivalent two-dimensional subalgebras of (\ref{sym2}). For reason of brevity, we only list the subalgebras without the 
corresponding conditions on the respective coefficients $a_i^j$:
\begin{align}\label{two}
\begin{split}
  &\langle \DDD, \vv_r \rangle, \\[.5ex]
  &\langle\DDD+a\vv_r, \vv_t \rangle,  \\[.5ex]
  &\langle\DDD+a\vv_x, \vv_t \rangle,  \\[.5ex]
  &\langle\DDD+a\vv_x, \vv_y \rangle,  \\[.5ex]
  &\langle\DDD+a\vv_r, \vv_\psi \rangle,  \\[.5ex]
  &\langle\DDD+a\vv_x, \vv_\psi \rangle, \\[.5ex]
  &\langle\vv_r+c\vv_\psi, \vv_t+ b\vv_\psi \rangle, \\[.5ex]
  &\langle\vv_r+c\vv_t, \vv_\psi \rangle, \\[.5ex]
  &\langle\vv_t+a\vv_x+c\vv_\psi, \vv_y+b\vv_\psi \rangle, \\[.5ex]
  &\langle\vv_t+a\vv_x, \vv_\psi \rangle, \\[.5ex]
  &\langle \vv_x+c\vv_\psi, \vv_y+b\vv_\psi \rangle, \\[.5ex]
  &\langle \vv_x, \vv_\psi \rangle,
\end{split}
\end{align}
where $c\in\{-1,0,1\}$ and $a,b \in \mathbb{R}$. Moreover, in the case $c=0$ we can scale the coefficient $b$ to obtain $b \in \{-1,0,1\}$. 
Additionally, if $c=b=0$ in the ninth subalgebra then we can set $a\in\{-1,0,1\}$.

\section{Group-invariant solutions}

Selected group-invariant solutions of (\ref{vort}) have been studied in \cite{huan04Ay} but without reference to classes of inequivalent 
subalgebras and without noting that it is possible to set $\beta = 0$. Consequently, some of the ans\"atze presented in \cite{huan04Ay}, which 
lead to a reduction of the number of independent variables of (\ref{vort}), are overly intricate. More precisely, they could be realised by 
means of considering reduction using one of the inequivalent subalgebras (\ref{one}) or (\ref{two}) and subsequently acting on the resulting 
invariant solutions by finite symmetry transformations.

We investigate potentially interesting group-invariant reductions upon using ans\"atze that are based on the above classification of 
subalgebras. If possible, we relate them to the solutions given in~\cite{huan04Ay}.

\subsection{Reductions with one-dimensional subalgebras}

Here we give the complete list of reduced equations of (\ref{vort}) with the parameters $F\ne 0$, $\beta=0$ based on the classification of 
inequivalent subalgebras (\ref{one}). In what follows $v$ and $w$ are functions of $p$ and $q$. 

\noindent {\bf 1.} $\langle \DDD+a\vv_r \rangle$. Suitable invariants of this subalgebra for reduction are
\begin{align*}
    &p = \phantom{-}x\cos(a\ln t) + y\sin(a\ln t), \\
    &q = -x\sin(a\ln t) + y\cos(a\ln t), \\
    &v = t\psi.
\end{align*}
Using them as new variables, the vorticity equation (\ref{vort}) is reduced to:
\begin{align*}
    &w-a(qw_p-pw_q) -F(v-a(qv_p-pv_q)) + {} \\
    &\quad v_qw_p-v_pw_q = 0,\\
    &w = v_{pp}+v_{qq}.
\end{align*}

\noindent {\bf 2.} $\langle\DDD+a\vv_x \rangle$. Invariants of this subalgebra are
\begin{align*}
    &p = x-a\ln t, \\
    &q = y, \\
    &v = t\psi,
\end{align*}
which reduce (\ref{vort}) to
\begin{align*}
    &w + aw_p - F(v+av_p) - v_pw_q+v_qw_p = 0,\\
    &w = v_{pp}+v_{qq}.
\end{align*}

\noindent {\bf 3.} $\langle\vv_r \pm \vv_t + a\vv_\psi \rangle$. Defining $\varepsilon=\pm 1$, we find the following invariants:
\begin{align*}
    &p = \phantom{-}x\cos\varepsilon t + y\sin\varepsilon t, \\
    &q = -x\sin\varepsilon t + y\cos\varepsilon t, \\
    &v = \psi-\varepsilon a t.
\end{align*}
In these variables, (\ref{vort}) reads
\begin{align*}
    &\varepsilon(qw_p - pw_q) - \varepsilon F(qv_p-pv_q + a) + {} \\
    & \quad v_pw_q - v_qw_p = 0, \\
    &w = v_{pp} + v_{qq}.
\end{align*}    

\noindent {\bf 4.} $\langle\vv_r + c\vv_\psi \rangle$. Here we have the invariants
\begin{align*}
    &p = \sqrt{x^2+y^2}, \\
  &q = t, \\
  &v = \psi + c\arctan\frac{x}{y}.
\end{align*}
This gives the same ansatz as was used in \cite{huan04Ay} (Case 2) upon additionally applying transformation (\ref{trans}). The corresponding 
reduced equation reads
\begin{align*}
    &w_q -Fv_q -\frac{c}{p}w_p = 0 \\
    &w=v_{pp}+\frac{1}{p}v_p.
\end{align*}
 
\noindent {\bf 5.} $\langle\vv_t + a\vv_x + c\vv_\psi \rangle$. Invariants of this subalgebra are
\begin{align*}
  &p = x - at, \\
  &q = y, \\
  &v = \psi - ct.
\end{align*}
This is an ansatz for a traveling wave solution in $x$-direction, which includes the well-known Rossby waves. In \cite{huan04Ay} a similar ansatz was combined with a traveling wave ansatz also in $y$-direction (Case~3). However, as was indicated above, the additional consideration of waves in $y$-direction is not necessary at this stage of reduction. The equation corresponding to the above ansatz is
\begin{align*}
    &aw_p - F(av_p-c)-v_pw_q+v_qw_p = 0, \\
    &w=v_{pp}+v_{qq}.
\end{align*}

\noindent {\bf 6.} $\langle\vv_x + c\vv_\psi \rangle$. A suitable ansatz for the invariants of this subalgebra is provided by
\begin{align*}
    &p = x,\\
    &q = y,\\
    &v = \psi - ct.
\end{align*}
In this case (\ref{vort}) is reduced to
\begin{align*}
    &Fc - v_pw_q + v_qw_p = 0, \\
    &w = v_{pp} + v_{qq}.
\end{align*}

\noindent {\bf 7.} $\langle\vv_\psi \rangle$. It is not possible to make an ansatz for $\psi$ for this subalgebra in the framework of the 
classical Lie approach. Hence, no reduction can be achieved through the gauging operator~$\vv_\psi$.

\subsection{Reductions with two-dimensional subalgebras}

To give also an example for a reduction using a two-dimensional subalgebra, let us consider the algebra $\langle\vv_t + a\vv_x + c\vv_\psi, 
\vv_y + b\vv_\psi\rangle$. In what follows $v$ is a function of $p$. The invariants of this algebra are
\begin{align}\label{ans2dim}
\begin{split}
    p &= x - at, \\
    v &= \psi - by- \frac{c}{a}x,
\end{split} 
\end{align}
provided that $a\ne 0$. The corresponding reduced ODE of (\ref{vort}) then reads
\[
    (a+b)v_{ppp} - Fav_p = 0
\]
with the general solution
\begin{align*}
    v &= v_1\exp\left(\sqrt{\frac{Fa}{a+b}}p\right)+ {} \\
    & {} v_2\exp\left(-\sqrt{\frac{Fa}{a+b}}p\right) + v_3,
\end{align*}
where $v_i=\const$, $i=1,2,3$. Transforming back to the original variables, renaming the constants $v_i$ and applying the transformation 
(\ref{trans}), we obtain the invariant solution
\begin{align*}
    \psi &= \psi_3 + \left(b+\frac{\beta}{F}\right)y+\frac{c}{a}\left(x+\frac{\beta}{F}t\right) + {} \\
    & {} \psi_1\exp\left(\sqrt{\frac{Fa}{a+b}}\left(x+\frac{\beta}{F}t-at\right)\right)+ {} \\
    & {} \psi_2\exp\left(-\sqrt{\frac{Fa}{a+b}}\left(x+\frac{\beta}{F}t-at\right)\right)
\end{align*}
which for $Fa / (a+b) < 0$ gives rise to a travelling wave solution.

For the singular case $a=0$, we cannot use ansatz (\ref{ans2dim}). Instead, we have the ansatz 
\begin{align*}
    p &= x,\\
    v &= \psi - ct-by
\end{align*}
and the reduced vorticity equation reads
\[
    Fc +  bv_{ppp} = 0
\]
which gives rise to a polynomial solution in the case $b\ne 0$. If $b=0$, we get the condition that $c=0$ and the ansatz $\psi=v(x)$ itself is 
the solution of (\ref{vort}).

\section{Summary and further\newline comments}

In the sections above, we discussed distinct cases of reduction by using inequivalent Lie subalgebras. The main advantages of this systematic approach are the following:
\begin{itemize}\itemsep=-0.5ex
    \item Simplification of equation (\ref{vort}) since we only have to consider the case $F\ne 0$, $\beta = 0$.
    \item Considering a minimal number of essential Lie subalgebras for reduction.
    \item Simplifying the ansatz for the reduced equations.
    \item Optimal preparation of the reduced equations for further investigations.
\end{itemize}
Upon using this approach, we have discussed all possible reductions by means of one-dimensional Lie subalgebras. We have to note that it is not possible to use the subalgebra $\langle \vv_\psi \rangle$ for obtaining group-invariant solutions, since in this case there is no way to make an ansatz for $\psi$. In principle, the remaining two-dimensional Lie subalgebras can be used for reduction as well.

Moreover, the differential equations obtained by reduction could again be investigated by means of symmetry techniques. In general, some of the symmetries of these reduced equations will be induced by the symmetries of the original equation. However, sometimes there may be additional symmetries that are not induced in this way \cite{kapi78Ay,olve86Ay} and which are called hidden symmetries \cite{abra06Ay}. They usually play an important role in the study of differential equations, as they may allow to reduce equations further than initially expected.

We may also note that it is still possible to generalise some of the results of this paper. By considering eqn.\ (\ref{vort}) as a system of two 
PDEs in the two dependent variables $\psi$ and $\zeta$, it is possible to construct partially invariant solutions \cite{ovsi82Ay}. For this 
class of exact solutions, one needs at least two dependent variables. For the first set of these dependent variables, it is possible to 
introduce new invariant variables, for the second set we keep the old noninvariant variables. The resulting reduced equations then also split in 
two sets of equations which have to be solved one after another. For this purpose, we could e.g.\ use subalgebras containing the operator 
$\vv_\psi$. In this case, it is still not possible to make an ansatz for $\psi$ but it is possible to do so for $\zeta$. The resulting reduced 
system of differential equations may give rise to a much wider class of exact solutions than pure group-invariant solutions. An investigation of 
this class of solutions for the case $F=0$, $\beta \ne 0$ will be given elsewhere.

\subsection*{Acknowledgments}

The research of this work was supported by the Austrian Science Fund (FWF), project P20632. AB is a recipient of a DOC-fellowship of the Austrian Academy of Science.

\end{multicols}

\end{document}